\documentclass[prl, reprint, twocolumn, superscriptaddress, longbibliography]{revtex4-2}

\usepackage{amsmath, amsthm, amssymb, amsfonts, braket, bm, empheq, array, lipsum}
\usepackage{dcolumn}
\usepackage{tabularx}
\usepackage[]{graphicx}
\DeclareGraphicsExtensions{.eps,.jpg,.png,.pdf}
\usepackage{hyperref, url}
\hypersetup{
colorlinks=true,
linkcolor=red,
filecolor=blue,
urlcolor=magenta,
}

\usepackage{dblfloatfix}
\usepackage[normalem]{ulem}
\UseRawInputEncoding

\newcommand{\be}{\begin{equation}} 
\newcommand{\ee}{\end{equation}} 
\newcommand{\bea}{\begin{eqnarray}} 
\newcommand{\eea}{\end{eqnarray}}

\newcommand{\Pe}{\mathrm{Pe}}

\newcommand{\rc}{\rho_{\mathrm{c}}}

\DeclarePairedDelimiter{\avg}{\langle}{\rangle}
\newcommand{\lD}{\mathbb{D}}
\newcommand{\vect}[1]{\bm{#1}}
\newcommand{\ii}{\mathrm{i}}

\newcommand{\eec}{\mathrm{e}}

\begin{document}
\title{Active model B$+$ from run-and-tumble particles on a triangular lattice}
\author{Abir Bhowmick}
\thanks{ab23rs019@iiserkol.ac.in}
\affiliation{Department of Physical Sciences, Indian Institute of Science Education and Research Kolkata, Mohanpur, 741246, India.}
\author{P. K. Mohanty}
\thanks{pkmohanty@iiserkol.ac.in}
\affiliation{Department of Physical Sciences, Indian Institute of Science Education and Research Kolkata, Mohanpur, 741246, India.}
\affiliation{Dutch Institute for Emergent Phenomena (DIEP), University of Amsterdam}
\date{\today}

\begin{abstract}
Active model B$+$ (AMB$+$) explains how motility-induced phase separation can be arrested by reverse Ostwald ripening, but its coefficients have not been derived from a microscopic lattice model that shows it. We provide an exact hydrodynamic description of a six-polarity lattice gas on a hexagonal geometry and employ a full gradient expansion with angular-harmonic elimination to rigorously derive AMB$+$ with every coefficient fixed by the P\'eclet number $\Pe$. Near the critical point $(3/4, 4)$, coarsening is ordinary; further away, field theory predicts reverse Ostwald ripening of vapor bubbles above $\mathrm{Pe}> 4\sqrt{3}$, and simulations show long-lived bubbly states.
\end{abstract}

\maketitle

\textit{Introduction.---} Self-propulsion is a defining feature of diverse active systems~\cite{Ramaswamy2017}, spanning fish schools~\cite{Katz2011} and bird flocks~\cite{Ballerini2008} to active colloids~\cite{Marchetti2016, Farage2015, Liebchen2017, Illien2017, vanderLinden2019, Zttl2023, Melio2025}, cytoskeletal structures~\cite{Bendix2008, Sanchez2012, Alvarado2013, Geisterfer2021}, molecular motors~\cite{Jlicher1997, Ndlec1997, Kruse2004, Liverpool2003}, and actin filaments~\cite{Schaller2010}. A central emergent phenomenon in these non-equilibrium systems is motility-induced phase separation (MIPS), wherein purely repulsive self-propelled entities spontaneously segregate into coexisting dense and dilute phases~\cite{Tailleur2008, Fily2012, Marchetti2013, Bialke2013, Redner2013, Cates2015}.

At large scales, MIPS is described by scalar field theories,  starting with active model B (AMB)~\cite{Wittkowski2014, Cates2013, Stenhammar2013, Speck2014, Solon2015}. Particle simulations, however, revealed features AMB cannot produce: vapor bubbles inside the dense phase that neither coarsen nor disappear~\cite{Stenhammar2014dim, Caporusso2020, Shi2020}. Active model B$+$ (AMB$+$) accounts for them by adding the lowest-order terms that make the current non-gradient~\cite{Nardini2017, Tjhung2018}, 
\begin{align}
    &\partial_t \phi =
    \nabla^2\left[\frac{\delta\mathcal{F}}{\delta\phi}+ \lambda|\bm\nabla\phi|^2\right] 
    - \zeta\, 
    \bm\nabla \cdot \left[(\nabla^2\phi)\bm\nabla\phi\right]
    \label{eq:AMB+}
\end{align}
with the baseline free-energy functional in terms of the scalar field $\phi\equiv \phi(\mathbf r,t)$, is $\mathcal{F}[\phi]=\int d\mathbf{r} \left[\frac r2\phi^2 +\frac u4\phi^4 + \frac\kappa2|\nabla\phi|^2\right]$, corresponding to the usual passive model B \cite{Hohenberg1977}. The scalar term $\lambda|\bm\nabla\phi|^2$ and the active vector stress $\zeta\,\bm \nabla \cdot \left[(\nabla^2\phi)\bm\nabla\phi\right]$ explicitly break time-reversal symmetry and detailed balance at the field level~\cite{Nardini2017, SolonPRE2018}. Crucially, the non-integrable term $\zeta$ generates sustained circulating currents in steady state, suppressing macroscopic coarsening and stabilizing microphase-separated patterns~\cite{Tjhung2018}.

This field-theoretic picture rests on coefficients that are normally treated as free parameters. Deriving them from particle dynamics is hard, because the non-gradient term appears only at fourth order in gradients. Fourth-order expansions have now been carried out for active Brownian particles (ABP)~\cite{ Rapp2019, Vrugt2023, PintoGoldbergSoto2026}. In the most recent of these, the two active coefficients come out with opposite signs, which predicts ordinary coarsening~\cite{PintoGoldbergSoto2026}. Active lattice gas systems provide a useful microscopic framework for investigating nonequilibrium collective phenomena, including MIPS, critical behavior, and transport in persistent active particles. Their dynamics have been studied extensively using analytical approaches and
numerical simulations~\cite{Thompson2011, Dandekar2020, Dittrich2021, Ray2024, Bhowmick2025}. A further advantage of these models is that, for appropriately constructed microscopic dynamics and a suitable macroscopic scaling, their evolution can be related directly to continuum equations. In particular, exact hydrodynamic limits have been derived for active lattice gases exhibiting MIPS and collective motion~\cite{Kourbane2018, Mason2023}. Building on this framework, subsequent studies have investigated typical fluctuations and large deviations, as well as hydrodynamic descriptions and critical behavior in other active lattice gas
models~\cite{Agranov2021, Agranov2023, Neville2025, Mukherjee2025, Erignoux2024}.
Despite this progress, the derivation of continuum theories from microscopic dynamics has so far yielded AMB field theory near criticality in one dimension~\cite{Neville2025}, leaving the corresponding derivation of AMB$+$ in two dimensions an open challenge.

In this Letter, we establish an explicit analytical connection between a microscopic lattice model of run-and-tumble particles (RTPs) and the two-dimensional continuum field theory of AMB$+$, providing a microscopic foundation for its phenomenological description. Our results are threefold. (i) Six-fold symmetry is the minimal lattice symmetry for which the fourth-order hydrodynamics is isotropic and can therefore take the AMB$+$ form. This requirement is the active-matter counterpart of the symmetry considerations that motivate the use  of triangular lattice in  cellular automata  of  Navier-Stokes equation ~\cite{FHP1986}. (ii) The critical point and all AMB$+$ coefficients are fixed by a unique combination of microscopic rates, namely the P\'eclet number $\Pe,$ which is dimensionless. Near criticality $\lambda<0$ (while $\zeta>0~\forall~\Pe>0)$ independently of the tumbling kernel, so the critical region shows ordinary MIPS, consistent with Ising criticality~\cite{Partridge2019, Dittrich2021, Saha2024, Bhowmick2025}. (iii) Away from criticality, a mean-field calculation shows that vapor bubbles ripen in reverse when $\Pe$ is increased above a threshold $\Pe^*=4\sqrt{3}$. There, simulations show reverse ripening of bubble pairs, arrest of coarsening length $L(t)$, and long-lived bubbly states.

\begin{figure}
    \centering
    \includegraphics[width=\linewidth]{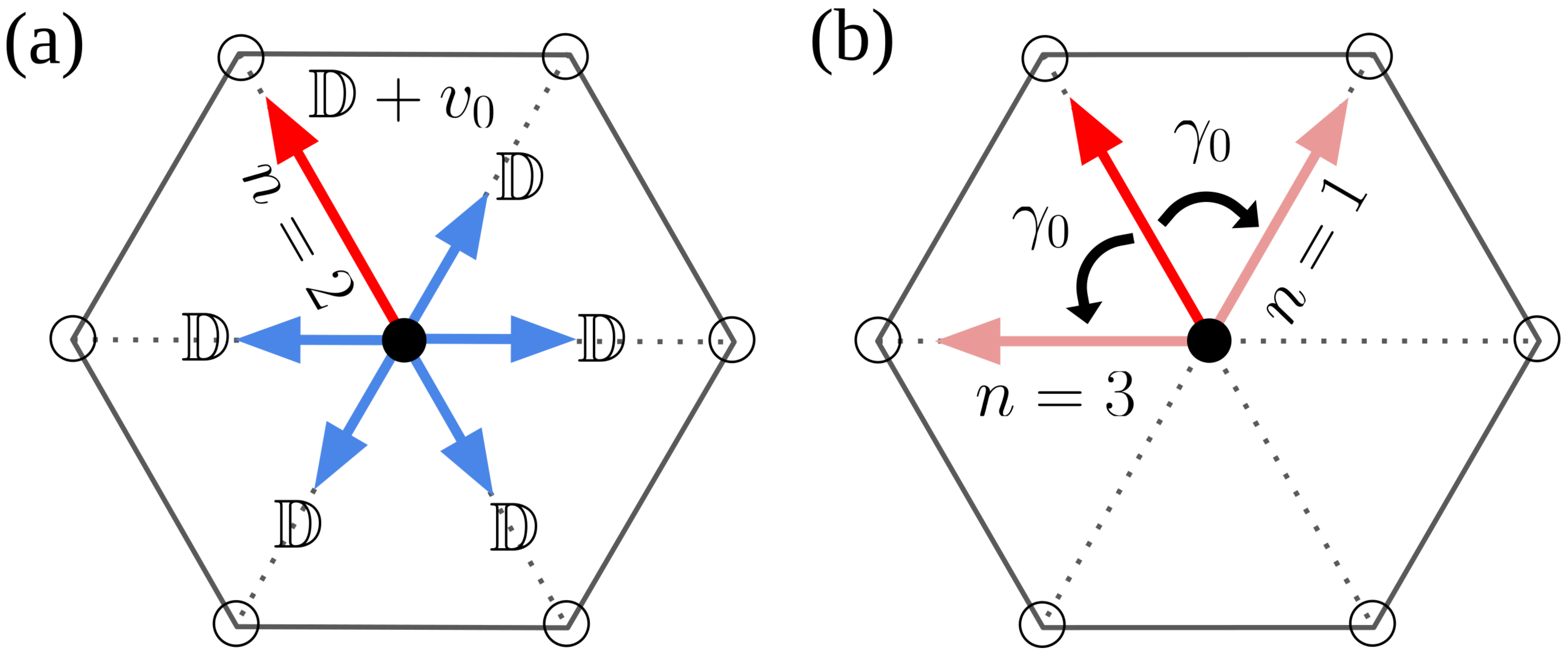}
    \caption{Dynamics of a  particle  having polarity $n$  (here $n=2$) is  shown on a hexagonal unit cell. (a) It attempt to move along $\vect{e}_{n=2}=\left(\cos\frac{2\pi}{3},\sin\frac{2\pi}{3}\right)$ with rate $\mathbb{D}+v_0$   or any of the  other $\vect{e}_{m\ne2}$ directions  each with rate $\mathbb{D}$, provided the target site is empty. (b) It can also change  its polarity  to $n-1$  or $n+1$ (pointing to $\vect{e}_{1}$  or $\vect{e}_{3}$  here), with rate $\gamma_0.$}
    \label{fig:shcematic_dynamics}
\end{figure}

\textit{Model and exact hydrodynamics.---} We consider $N$ particles on a triangular lattice of $L\times L$ sites with single occupancy. A particle carries one of six polarities $n=0,\dots,5$, with unit vectors $\vect{e}_n=\left(\cos\frac{n\pi}{3},\sin\frac{n\pi}{3}\right)$. The particle can: (i) diffuse to any of its empty nearest neighboring sites with rate $\lD$, (ii) hop in along its polarity direction $\mathbf{e}_n$, with rate $v_0$ if it is vacant, (iii) tumble to any of the two nearest neighboring directions $n\pm1$ with rate $\gamma_0$. We take the weakly active diffusive limit~\cite{Kourbane2018} $v_0=v/L$, $\gamma_0=\gamma/L^2$ with $\lD$  remaining ${\cal O}(1)$, where  the hydrodynamic equations obtained from the evolution equation of the microscopic occupation variable $\sigma_{\bm i}^n(h)\in\{0,1\}$, at site $\bm i$ and time step $h$,  are exact (see~\cite{SM} for detailed derivation).

\textit{Continuum Limit and Scale Invariance.---} This diffusive rescaling of space $\vect{r}=\bm{i}/L$ and time $t=\tau/L^2$, makes all three processes (diffusion, self-propulsion and tumbling) contribute equally in the hydrodynamic limit ($a\to0$ or $L\to\infty$). In this large $L$ limit, local correlations become negligible at the macroscopic
scale, allowing the evolution of the coarse-grained density $\rho^n(\vect r,t)$ to be described by a closed hydrodynamic equation. This description relies on the \emph{``local equilibrium''} assumption~\cite{spohn2012}, under which the local statistics are approximated by product Bernoulli measures parametrized by the local densities. The resulting correlation-free evolution equation reads
\begin{align}
    \partial_t\rho^n
    &=D\left[(1-\rho)\nabla^2\rho^n+\rho^n\nabla^2\rho\right]
    -v\,\mathbf e_n\cdot\bm\nabla[(1-\rho)\rho^n]\nonumber\\
    &\quad+\,\gamma\,(\rho^{n-1}+\rho^{n+1}-2\rho^n),
    \label{eq:hydro}
\end{align}
where $D=\frac32\lD$ is the continuum diffusion constant, and $\rho=\sum_n\rho^n$ is the total density.

To analyze the evolution dynamics of Eq.~\eqref{eq:hydro}, we reduce it to a dimensionless form using $\bm r\to\sqrt{D/\gamma}$ and $t\to t/\gamma$, and in this unit the macroscopic equation reads
\begin{align}
    \partial_t\rho^n
    &= (1-\rho)\nabla^2\rho^n + \rho^n\nabla^2\rho
    - \Pe\,\vect e_n\cdot\bm\nabla[(1-\rho)\rho^n]\nonumber\\
    &\quad+\,\rho^{n-1}+\rho^{n+1}-2\rho^n,
    \label{eq:hydro_discrete}
\end{align}
$\Pe=v/\sqrt{D\gamma}$ being P\'eclet number. This form matches with the noiseless case of the $d$-dimension generalization form of the hydrodynamic equation derived in Ref~\cite{Mukherjee2025}. 

\begin{figure}
    \centering
    \includegraphics[width=.9\linewidth]{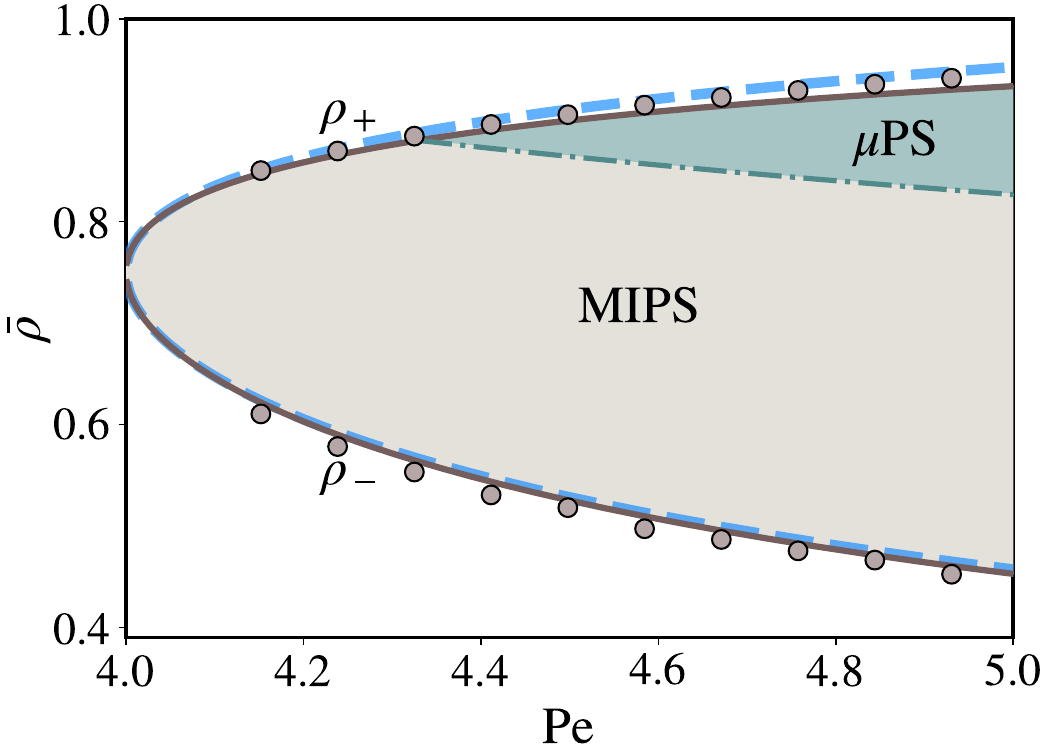}
    \caption{The binodal curve. Coexisting densities  $\rho_\pm$ extracted from microscopic lattice simulations (symbols) are compared with  the exact binodal curve (thick solid line) obtained by solving the hydrodynamic equations analytically; they meet at the critical point $\Pe_c=4,~ \rho_c=3/4$. The dashed line following the symbols is the binodal of the equivalent AMB$+$ field theory described by Eqs.~\eqref{eq:AMB+} and \eqref{eq:mapping}. Micro-phase separation occurs deep in the phase-separated region when $\Pe$ crosses a threshold given by Eq.~\eqref{eq:RoR}; this $\mu$PS region is shaded differently.}
    \label{fig:binodal}
\end{figure}

\textit{Angular harmonics and isotropy.---} We write the six densities in terms of discrete harmonics, $p_\ell=\sum_n e^{i\ell n\pi/3}\rho^n$, which obey $p_{\ell+6}=p_\ell$. The conserved density is denoted by $p_0=\rho$, the polarization by $p_1=m_x+\,\ii\,m_y$, and $p_2=Q_1+\,\ii\,Q_2$ denotes the nematic tensor, which retains the traceless second orientational moment, with $Q_1=M_{xx}-M_{yy}$ and $Q_2=2M_{xy}$, where $M_{ij}=\sum_n \eec_i^n\eec_j^n\rho^n$. The third harmonic $p_3$, describes the alternating-polarity mode. The other two harmonics are $p_4=p_2^*$ and $p_5=p_1^*$.

In terms of the harmonics $p_\ell$, Eq.~\eqref{eq:hydro_discrete} takes the form
\begin{align}
    \partial_t p_\ell
    &=w\nabla^2p_\ell+p_\ell\nabla^2\rho
    -\frac{\Pe}{2}\left[
    \partial_+(wp_{\ell-1})+\partial_-(wp_{\ell+1})
    \right]\nonumber\\
    &\quad-\Gamma_{\ell}p_\ell,
    \label{eq:harm}
\end{align}
where  $w=1-\rho$, $\Gamma_\ell=2[1-\cos(\ell\pi/3)]$, and $\partial_\pm \equiv \partial_x\pm\ii\partial_y$. Clearly, except the zeroth order $\rho= p_0$ which follow a conserved dynamics, all higher order harmonics are  non-conserved (as $\Gamma_\ell>0~\forall~\ell>0$) and relaxes fast to their steady state values. Thus we adiabatically eliminate these modes and perform a gradient expansion. 
By eliminating the  polarization $p_1 \sim \mathcal{O}(\nabla)$ and the second harmonic $p_2 \sim \mathcal{O}(\nabla^2),$   and keeping  terms up to $\mathcal{O}(\nabla^3)$, we obtain macroscopic continuity equation for $\rho\equiv\rho(\mathbf{r},t)$ (see details at~\cite{SM}):
\begin{eqnarray}
    &&\partial_t \rho = \nabla^2\left[
    \mu - \hat\kappa\,\nabla^2\rho
    + \hat\lambda\,|\bm\nabla\rho|^2 \right] - \hat\zeta\,\bm\nabla\cdot\left[ (\nabla^2\rho)\bm\nabla\rho \right], \cr
    &&\mu(\rho) = \rho + \frac{\Pe^2}{2}
    \left(\rho - \frac{3}{2}\rho^2
    + \frac{2}{3}\rho^3\right), \cr
    &&\hat\kappa(\rho) = \frac{\Pe^2}{24}
    (1-\rho)^2(2\rho-1)\left[12 + (1-\rho)\Pe^2 \right], \cr
    &&\hat\lambda(\rho) = \frac{\Pe^2}{24}(1-\rho)^2
    \left[2\Pe^2(3\rho-2)-24\right], \cr
    &&\hat\zeta(\rho) = \frac{\Pe^2}{24}
    \Big[12(-3+9\rho-6\rho^2) \nonumber\\
    &&\qquad\qquad\qquad +\,3\Pe^2(1-\rho)^2(2\rho-1)\Big],
    \label{eq:density}
\end{eqnarray}

Equation~\eqref{eq:density} takes the structure of the AMB$+$ framework defined in Eq.~\eqref{eq:AMB+}, but the coefficients are nonlinear functions of the field $\rho$. The explicit retention of the second harmonic $p_2$ is crucial in obtaining this form; physically, this mode represents transient nematic/hexatic ordering at the interfaces of dense clusters. Because $p_2$ feeds back into $p_1$, it alters the active currents along the phase boundaries, generating an exact $\mathcal{O}(\Pe^4)$ correction to the density current, which we find essential for formation and stability of vapor bubbles in  (micro-phases).

Note that we ignore contribution from $p_3 \sim \mathcal{O}(\nabla^3)$ in the gradient expansion as it contributes $\mathcal{O}(\nabla^5)$  term to  Eq.~\eqref{eq:density}, which is irrelevant~\cite{Fejos2026} in two dimensions. Also on a square lattice, there are only four harmonics obeying $p_\ell=p_{\ell+4}$ and the identification $p_0=p_4 \sim \mathcal{O}(\nabla^4)$ generates an anisotropic $4$th order gradient. Thus the six-fold symmetry is the minimal lattice symmetry for an isotropic AMB$+$ description, just as it is for the Navier-Stokes limit of lattice gas automata~\cite{FHP1986}.

This hydrodynamic equation provides us with the full coarse-grained free energy $f(\rho)=\int\mu(\rho)\,d\rho$, which depends only on a single parameter $\Pe$. The homogeneous state $\rho(\mathbf{r},t)=\bar \rho$ becomes unstable when the local compressibility $\mu'(\bar\rho)$ vanishes, together with $\mu''(\bar\rho)$; this locates the critical point exactly at $\rho_c=3/4$ and $\Pe_c=4$; the system phase separates when the P\'eclet number $\Pe>\Pe_c.$

The boundary between the homogeneous (mixed) fluid and the motility-induced phase-separated (MIPS) state is set by the binodal lines which can be calculated using the pseudo-density and the pseudo-potential as suggested in Refs.~\cite{SolonPRE2018, SolonNJP2018}. Far from criticality, where the gradient expansion is not controlled, we instead obtain the binodals from steady flat-interface solutions of the full hydrodynamics, Eq.~\eqref{eq:hydro} (see~\cite{SM} for details). Figure~\ref{fig:binodal} compares them with  the Monte Carlo  simulations  on the lattice. The agreement is excellent across the coexistence region. Thus, the coarse-grained hydrodynamic equations captures the large scale features of the microscopic model quite well.

\textit{Field theory near criticality.---} Expanding $\rho=\rho_c+\phi$ about the critical density $\rho_c=3/4$, recasts the full hydrodynamic equation to AMB$+$ form defined in Eq.~\eqref{eq:AMB+} with constant coefficients, tunable by the dimensionless propulsion parameter $\Pe$, 
\begin{eqnarray}
    &&r=\mu'(\rho_c)=1-\frac{\mathrm{Pe}^2}{16},\cr 
    &&u= \frac{\mu'''(\rho_c)}6=\frac{\mathrm{Pe}^2}{3},\cr
    &&\kappa=\frac{\mathrm{Pe}^2(\mathrm{Pe}^2+48)}{3072},\cr
    &&\lambda=\frac{\mathrm{Pe}^2(\mathrm{Pe}^2-48)}{768},\cr
    &&\zeta=12\,\kappa.
    \label{eq:mapping}
\end{eqnarray}

\begin{figure}
    \hspace*{-.2in}
    \includegraphics[width=.95\linewidth]{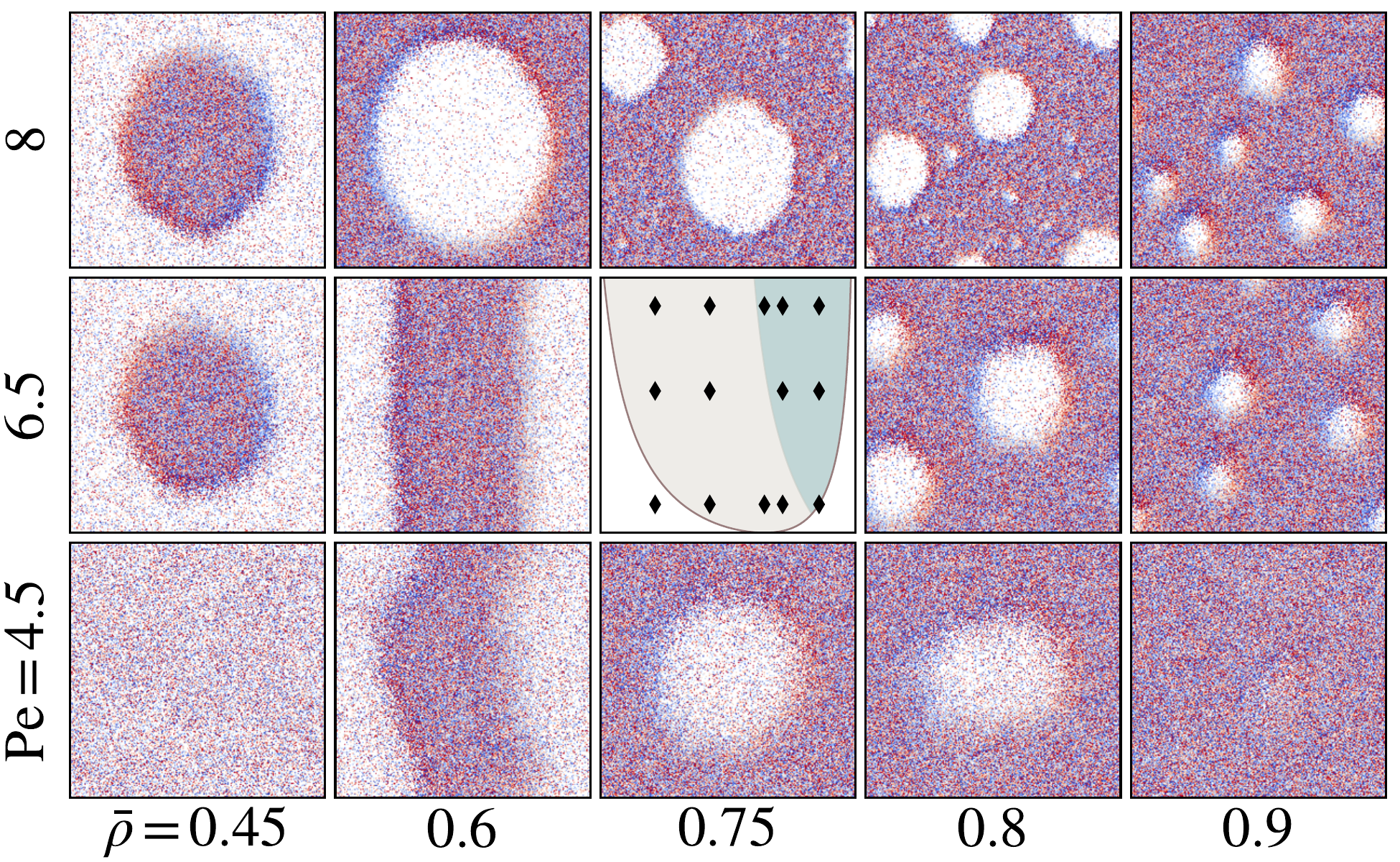}
    \caption{Microscopic simulation snapshots of the system at different points in the $\rho_0$-$\text{Pe}$  plane shown in the central inset,  where regions of MIPS and $\mu$PS  are shaded differently. In the MIPS state, the system exhibits bulk domains, in contrast to the $\mu$PS state where finite-size vapor bubbles are stabilized. The snapshots are taken at $10^6$ MCs and $L=256$.}
    \label{fig:phase_diagram}
\end{figure}

Equation~\eqref{eq:mapping} is the central result of this Letter: it provides an explicit dictionary between the microscopic six-polarity RTPs in 2D and the AMB$+$ field theory. Unlike in the AMB$+$ phenomenology where the parameters $\{r, u, \kappa, \lambda,~\zeta\}$, are independent control parameters, here they are fixed by the microscopic transition rates through $\Pe$. This mapping provides a direct route from microscopic parameters ($\Pe,~\bar\rho$) to macroscopic phase behavior. 

\textit{Microphase separation.---} The characteristic feature of  AMB$+$ is formation of finite size clusters  or bubbles~\cite{Tjhung2018}.
Deep in the phase-separated state, a dense system creates steady vapor bubbles with interfaces spanning densities from $\rho_+$ to $\rho_-$, balancing the pressure $P(\rho_+)=P(\rho_-)+\frac{\sigma}{R}$, where $\rho_-$ is the bulk density inside the droplet and $\rho_+$ is the same outside the droplet, and $\sigma$ is the pseudo-tension. A positive tension gives the ordinary Gibbs-Thomson shift, whereas when $\sigma<0$, small bubbles are likely to sit in denser fluid in comparison to large ones, allowing particles to diffuse from small to large bubbles, resulting in growth of small bubbles in expense of large ones. Within the established framework of AMB$+$ field theory~\cite{Tjhung2018}, this reverse Ostwald ripening (ROR) process is governed by the competition between the two nonequilibrium active gradient terms $\lambda$ and $\zeta$. In particular, $\sigma$ becomes negative in the regions  $\lambda,~\zeta$ have the same sign, favoring ROR. Our microscopic-to-continuum mapping shows that [from Eq.~\eqref{eq:mapping}], in the near-critical regime $4<\Pe<4\sqrt{3}$, the system favors forward Ostwald ripening ($\lambda,~\zeta$ having opposite signs) and remains driven towards macroscopic phase separation (MIPS). On the other hand, increased activity beyond $\Pe>4\sqrt{3}$ reverses the process.

For a generic density $\bar\rho \ne \rho_c$, a simple estimate of the microphase-separated ($\mu$PS) region inside the binodal can be obtained  from Eq.~\eqref{eq:density} by expanding $\hat\lambda(\rho),~\hat\zeta(\rho)$ near $\rho= \bar\rho$. Since $\hat\zeta(\bar\rho)>0$ for any $\bar\rho$, the sign of $\hat\lambda(\bar\rho)$ sets the criteria for  $\mu$PS as,
\begin{equation}
    \Pe>\Pe^*(\bar\rho)
    =
    \frac{2}{\sqrt{\bar\rho-2/3}} .
    \label{eq:RoR}
\end{equation}
At $\bar\rho=\rc$ this gives $\Pe^*=4\sqrt3\simeq6.93$. However, one must keep in mind that Eq.~\eqref{eq:RoR} is only an estimate, not a controlled result, since it evaluates $\hat\lambda(\rho)$ at a single reference density while the interface spans the whole coexistence range. Physically, this threshold indicates the point where the active run length sufficiently dominates over local lattice constraints, allowing interfacial active fluxes to stabilize finite cluster sizes.

The complete phase diagram is presented in Fig.~\ref{fig:binodal}. The $(\Pe, \bar\rho)$ plane is divided into three distinct regions: homogeneous fluid region, motility-induced phase-separated region, and microphase separation region ($\mu$PS). The boundary between the homogeneous (mixed) fluid and the MIPS state is separated by the binodals (solid line), whereas $\mu$PS occurs within the phase-separated region when $\Pe$ exceeds $\Pe^*$ (dashed dot line). We also present in Fig.~\ref{fig:phase_diagram} some steady-state snapshots of the system in different regions marked in the phase diagram. Clearly, the steady-state snapshots mark the transition line defined by Eq.~\eqref{eq:RoR}, capturing the onset of ROR quite well. 

\begin{figure}[t]
    \centering
    \includegraphics[width=\linewidth]{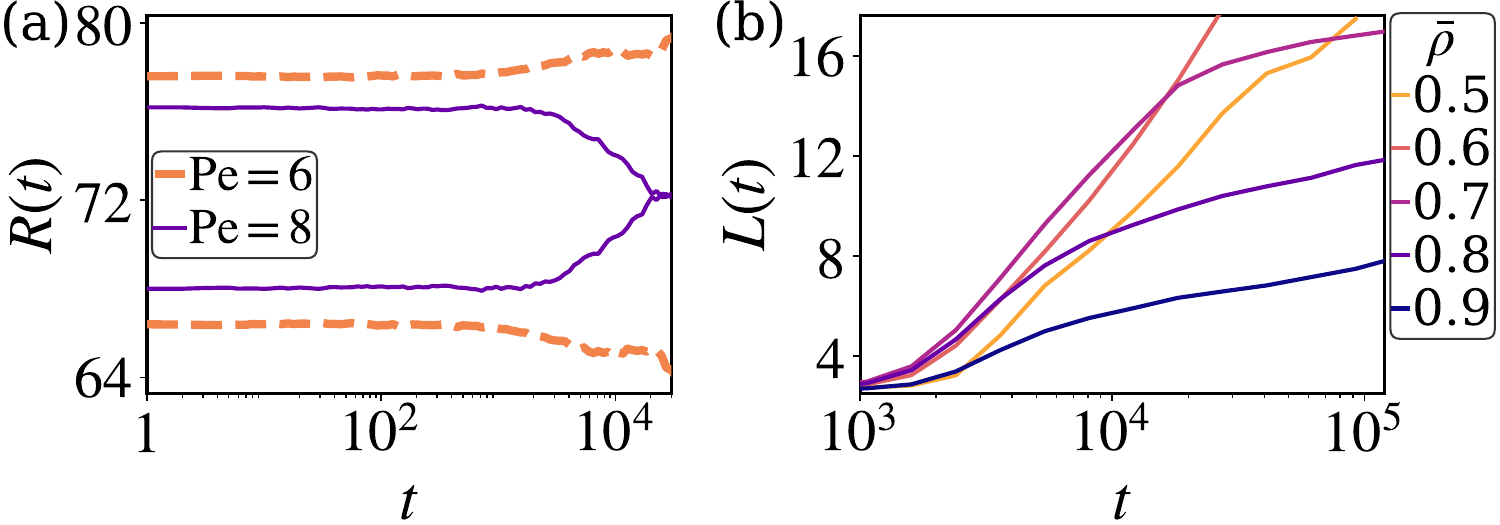}
    \caption{(a) Evolution of low-density bubbles in $2L\times L$ system. Initially, two circular voids of radius $R_1>R_2$ are created at the center of each $L\times L$-square ($L=256$).  (b)  As expected from \eqref{eq:RoR} for $\Pe=20$, growth of coarsening length $L(t)$ is arrested in $\mu$PS  state ($\bar\rho=0.7,~0.8,~0.9$), whereas usual  LS scaling is observed for $\bar\rho=0.5,~0.6$. Here, the system size considered for simulation is $L= 512$. }
    \label{fig:RoR}
\end{figure}

\textit{Simulations.---} To test the ROR process directly, we prepare an $2L\times L$ system ($L=256$) with two circular vapor bubbles of radii $R_1>R_2$, one in each $L\times L$ half, at $\bar\rho=0.75$ [Fig.~\ref{fig:RoR}(a)]. At $\Pe=6.0$, the larger bubble grows at the expense of the smaller one, as in ordinary Ostwald ripening. At $\Pe=8.0$ however, the trend reverses: the larger bubble shrinks, and the smaller one grows. Both observations agree with Eq.~\eqref{eq:RoR}. To follow the coarsening behavior, we take a reference $\Pe=20.0$, and observed the growth of the coarsening length scale $L(t)=2\pi/\avg{q}$ where $\avg{q}=\int q\,S(q,t)\,dq/\int S(q,t)\,dq$ [Fig.~\ref{fig:RoR}(b)]. For $\bar\rho=0.5$ and $0.6$, $L(t)$ grows  until they hit the bounds set by the size of the system; the growth follows $t^{1/3}$ law~\cite{LifshitzSlyozov1961, Huse1986} with the logarithmic corrections expected for conserved dynamics~\cite{Bhowmick2026} (see supplementary material~\cite{SM} for details). For $\bar\rho\ge0.7$, above the estimate $\bar\rho^*=0.677$ from Eq.~\eqref{eq:RoR}, the minority phase consists of vapor bubbles and $L(t)$ saturates at a finite value.

\textit{Discussion.---} We have derived AMB$+$ from a run-and-tumble lattice gas whose hydrodynamics is exact. The derivation fixes every coefficient through the P\'eclet number $\Pe$, which directly relates to the microscopic transition rates. The information about the lattice structure enters the hydrodynamic description only through the ratio $\Gamma_2/\Gamma_1$, where $\Gamma_{1,2}$ are the relaxation rates of the polarization and nematic modes, respectively. If microscopic tumbling kernel allows all possible turns  $(\pm\pi/3,\pm 2\pi/3,\pi)$ with rates $(\gamma_1,\gamma_2, \gamma_3)$ respectively, we have $\Gamma_2/\Gamma_1= 3(\gamma_1 + \gamma_2)/ (\gamma_1+3 \gamma_2 + 2 \gamma_3).$ This generates AMB$+$ phenomenology, with slightly modified coefficients. The fact that the homogeneous state near criticality destabilizes to form  MIPS (not $\mu$PS) is generic, independent of the tumbling kernel. 

Three conclusions follow. First, a six-fold lattice is the minimal choice that gives an isotropic AMB$+$ at fourth order, whereas a square-lattice model is already anisotropic at the order where $\zeta$ appears. Second, the critical region shows ordinary coarsening, consistent with Ising criticality~\cite{Partridge2019, Dittrich2021, Saha2024, Bhowmick2025}. Third, beyond the critical region, the exact hydrodynamics predicts, at mean-field level, reverse ripening of vapor bubbles for $ \Pe> \Pe^*= 2/\sqrt{ \bar \rho-2/3)}$, and simulations show reverse Ostwald ripening of bubble pairs and long-lived bubbly states at high density.

These results give a microscopic route to the regime of AMB$+$ that has so far been explored only phenomenologically~\cite{Tjhung2018, Shi2020}.
Our approach generalizes directly to other lattice symmetries, to mixtures with nonreciprocal couplings, and to fluctuating hydrodynamics, where the same expansion should yield the AMB$+$ noise~\cite{Agranov2021}. Given that dense active fluids naturally exhibit local hexatic structure~\cite{Digregorio2018, Klamser2018}, the mechanism unveiled here provides a direct physical link between local packing geometry and the macroscopic stabilization of active microphases.

\begin{acknowledgments}
    \textit{Acknowledgement ---} PKM gratefully acknowledge the support, hospitality, and stimulating research environment provided by the Dutch Institute for Emergent Phenomena (DIEP) and the Faculty of Science (FNWI) at the University of Amsterdam (UvA) during my tenure as a Global South Fellow. 
\end{acknowledgments}

\bibliography{triRTP_arXiv.bbl}
\end{document}